\documentclass[11pt,twoside]{article}
\usepackage{CAGN2019}
\usepackage{graphicx}

\usepackage[T1]{fontenc} 

\usepackage{latexsym}
\usepackage{verbatim}

\usepackage{ifpdf}  
\ifpdf  
      \DeclareGraphicsExtensions{.pdf,.png,.jpg}  
\else  
      \DeclareGraphicsExtensions{.eps}  
\fi 

\begin{document}

\vskip 1.0cm
\markboth{Monique M. de Brito et al.}{A spectrophotometric study of planetary nebulae and HII regions in the M83 galaxy}
\pagestyle{myheadings}
%
%
\vspace*{0.5cm}
\parindent 0pt{Poster}


\vspace*{0.5cm}
\title{A spectrophotometric study of planetary nebulae and HII regions in the M83 galaxy}

\author{Monique M.~Brito$^1$ and O.~Cavichia$^1$}
\affil{$^1$Instituto de F\'isica e Qu\'imica, Universdidade Federal de Itajub\'a, Av. BPS, 1303, 37500-903, Itajub\'a-MG, Brazil.\\}

\begin{abstract}
Low and intermediate mass stars (0.8-8M$\odot$) in the end of their evolution pass through a series of events of mass loss, which contributes for the enrichment of the interstellar medium. The end of this evolutionary process is preceded by the planetary nebulae (PNe) phase, representing an important source of information for understanding the chemical enrichment of galaxies. The M83 barred galaxy is a relatively nearby galaxy, which allows the spectrophotometric study of its photoionized objects. In the literature, HII regions have been extensively explored in this galaxy, though its PNe population has not previously been addressed. In this work we report, for the first time, a spectrophotometric study of the PNe population in the M83 galaxy. The data was observed with the Gemini GMOS multi-object spectrograph in 2014 using two grating configurations (R400 and B600) and under an excellent seeing condition (< 0.8 arcsec). These data are curently being reduced with the Gemini Pyraf package. In the first phase of this project we wish to confirm spectroscopically the PNe candidates of the sample to further analyze, for the first time, the radial gradient of chemical abundances from the PNe population in this galaxy.


\end{abstract}

\section{Introduction}

The concept of stellar populations distinguishes objects belonging to a galaxy according to a series of parameters: age, chemical composition, spatial distribution and kinematic characteristics. The analysis of the different star populations
gives us a much information about the evolution of galaxies. As for spiral galaxies, most information on chemical abundances comes from photoionized nebulae, such as PNe and HII regions.

PNe and HII regions represent important links to the understanding the enrichment and chemical composition of galaxies. Both are considered as the best tracers of the chemical evolution of galaxies \citep{Denise}.

PNe are the result of the final stages of life of a star with mass in the range of 0.8-8M$\odot$, also called low mass and intermediate stars. The gas layers ejected from these stars represent a source of information to understanding the influence of PNe in the chemical enrichment of galaxies. The HII regions, so called by the amount of atomic and ionized hydrogen present in these clouds of gas, are regions of star formation. In spiral galaxies, HII regions are concentrated in the spiral arms and are very important in determining the chemical composition of galaxies.

The galaxy M83 (or NGC 5236), known as  Southern Pinwheel Galaxy, is a barred spiral (SAB (s)). M83 is located approximately 4.59 Mpc away from us in the direction of the Hydra constellation. In the literature there is a vast content regarding the HII regions in this galaxy on different subjects, including the determination of the radial gradient of oxygen abundance for M83 \citep{Bresolin}. However the population of PNe to date has not been studied in detail in this galaxy.

\citet{Herrmann} performed a photometric survey of PNe candidates in M83, where they obtained precise astrometric positions of these objects. However, a PN can be confused with HII regions and/or remnants of supernovae at these distances in photometric studies. The M83 galaxy, to date in the literature, does not have a spectroscopic survey regarding its PNe population. In view of this, it is necessary and interesting to confirm spectrophometrically whether these objects are actually PNe.

Our work aims a the spectrophotometric characterization of this sample of PNe in M83. In section 2 we present the procedures adopted, such as information regarding observation and data processing. In section 3 we discuss what the expected results are, since this is an undergoing project. And finally in section 4, our final considerations and perspectives.

\begin{figure}[h]
\begin{center}
\hspace{0.25cm}
\includegraphics[scale=0.6]{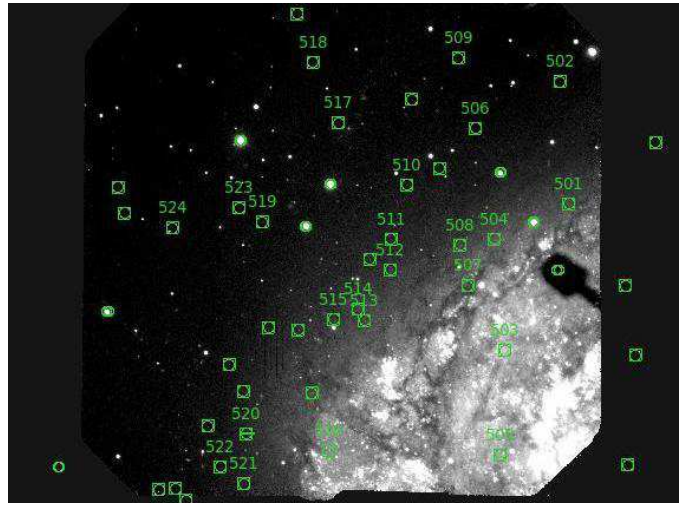}
\caption{Field of galaxy M83 that was observed. The positions of the candidates for PNe are represented by the green squares, and the circles represent stars that were used for aligning the mask.}
\label{m83}
\end{center}
\end{figure}

\section{Procedure}
\label{procedure}

The observations were made with the 8m Gemini South telescope, located in Chile in 2014. The multi-object spectrograph (GMOS) was used for these observations. This mode of spectroscopy was chosen because it was advantageous to simultaneously collect several spectra of objects located in the field of the telescope. 

\begin{figure}[h]  
\begin{center}
\hspace{0.25cm}
\includegraphics[scale=0.3]{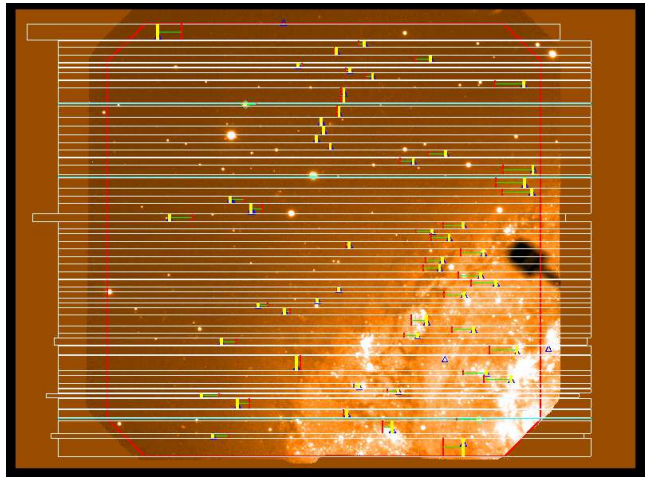}
\caption{One of the spectroscopic masks made, in this example for the B600 grating. The positions of the slitlets in the masks exactly match the positions of the PNe in which we wish to obtain their spectra.}
\label{fig1}
\end{center}
\end{figure}

Only one field of M83 was observed. The figure \ref{m83} shows the M83 field in which MOS spectroscopy was made and the observed PNe sample are indicated by green squares. Pre-images of the analyzed field of M83 were used to produce the spectroscopic masks (figure \ref{fig1}), with the purpose of selecting only our objects of interest, the candidates for PNe. We employed the R400 and B600 gratings, in which differ only in the adopted central wavelengths and spectral coverage. Our data were obtained under an excellent seeing condition (< 0.8 arcsec).

The data is currently being reduced with the Gemini PyRAF package, which allows scripts and execution of Image Reduction and Analysis Facility (IRAF) tasks from the Python environment. The reduction of data from this project is based on the tutorial available on the Gemini Observatory platform for multi-object spectra reductions. This tutorial is based on the work of \citet{TutorialGmos}.

The reduction process included bias and flat correction, twilight flats, mosaic the CCDs, wavelength calibration and flux calibration.

\section{Results}

This work is still under development, and in this first moment we intend to carry out the spectroscopic confirmation of the PNe sample of the M83 galaxy. Once the process of data reduction is finished, we will have the spectra of these objects and in this way we can characterize them.

The objects of our sample are all PNe candidates, which do not have spectroscopic observations available in the literature, but with enough quality and distance to derive physical parameters and possibly chemical abundances. 

At the conclusion of this work we hope to characterize the population of PNe as well as cataloging and discovery of new HII regions and/or supernovae remnants in this galaxy.  

\section{Conclusions}
\label{discussion}

In this work we present a study about the PNe population in the galaxy M83, observed with the Gemini multi-object spectrograph (GMOS). We intend to carry out the spectrophotometric characterization of these objects, as they do not have, up to the present moment in the literature, a detailed spectroscopic study of the PNe population in this galaxy. 

The spectroscopic confirmation of the sample of PNe in the galaxy M83 will later enable us to study of the chemical enrichment of the galaxy's interstellar medium. In this way, the radial
gradient of chemical abundances of the PNe population in M83 may be obtained for the first time.

These data will be fundamental to study the chemical evolution of the M83 galaxy, and may contribute to the study of the influence of bars on the radial gradient of chemical abundances of barred spiral galaxies \citep{Oscar}.

\acknowledgments We would like to thank FAPEMIG for the financial support. 

\bibliographystyle{aaabib}
\bibliography{brito}

\end{document}